\documentclass[twocolumn,prl,aps,showpacs,superscriptaddress]{revtex4}
\usepackage{xspace,amsmath,amsfonts,amsthm,amssymb,amsbsy,graphicx,color}

\begin{document}

\title{Engineering Giant Nonlinearities in Quantum Nano-Systems}

\author{Kurt Jacobs} 

\affiliation{Department of Physics, University of Massachusetts at Boston,
Boston, MA 02125, USA}

\affiliation{Hearne Institute for Theoretical Physics, Louisiana State
University, Baton Rouge, LA 70803, USA}

\author{Andrew J. Landahl} 

\affiliation{Sandia National Laboratories, Albuquerque, NM, 87185, USA}

\affiliation{Center for Advanced Studies, Department of Physics and
Astronomy, University of New Mexico, Albuquerque, NM, 87131, USA}

\begin{abstract} 
We describe a method to engineer giant nonlinearities in, and probes to measure  nonlinear observables of, mesoscopic quantum resonators. This involves tailoring the Hamiltonian of a simple auxiliary system perturbatively coupled to the resonator, and has the potential to engineer a wide range of nonlinearities to high accuracy. We give a number of explicit examples, including a readily realizable two-qubit auxiliary system that creates an $x^4$ potential and a $\chi^{(3)}$ (Kerr) nonlinearity, valid to fifth-order in the perturbative coupling. 
\end{abstract}


\maketitle 

In the last few years there has been rapid progress in the development of nano-mechanical resonators and superconducting microwave oscillators~\cite{LaHaye09, Majer07, Houck07, Regal08}. Both kinds of oscillators can be combined with other electrical elements to form mesoscopic devices, and because of their high frequencies, both have the potential to realize quantum behavior. If either kind could be built with nonlinearities that were strong in the quantum regime (so-called {\em giant} nonlinearities~\cite{Imamoglu97}), this would open up many potential applications. These include quantum computing~\cite{Lloyd99}, simulating many-body systems~\cite{Hartmann06}, and studying the quantum-to-classical transition~\cite{Bhattacharya00}. However, directly constructing resonators with such nonlinearities is not possible, at least with current technology~\cite{Woolley08}.  

Measuring nonlinear observables of resonators is also important; it reveals signatures of quantum dynamics~\cite{Martin07, Jacobs07b}, and we expect it to have future applications in feedback control and adaptive measurement~\cite{Wiseman95, Jacobs07c}. To do so one must find mesoscopic devices that have interactions with resonators that are proportional to these observables. Physically building such interactions is a very challenging task. The only method suggested to date that could generate a range of nonlinearities or nonlinear interactions requires active control of an auxiliary system on fast  time-scales~\cite{Jacobs07x, Jacobs08x}. Here we present a method to do this that does not suffer this limitation.    

We recall first that it is well-known in the field of quantum optics that a low-dimensional system can generate an effective nonlinearity in a larger system. For example, four-level atoms with the right level structures, when detuned appropriately from a cavity mode, will create $\chi^{(3)}$ and other nonlinearities for the mode~\cite{Imamoglu97, Schmidt96}. This is promising, because there are a number of low-dimensional mesoscopic systems that can be interfaced with resonators. However, to make this practical one must find systems with the right internal structure to generate the desired perturbation, and that can be readily realized. The simple and physically intuitive examples explored in quantum optics are natural for multi-level atoms, but not necessarily for mesoscopic systems. In addition, we wish to engineer nonlinear probes as well as nonlinarities. To achieve this goal we introduce a systematic method for finding auxiliary systems to generate a given dynamics, including calculating this dynamics to high-order in the perturbation. This allows us to search over many auxiliary systems to find the simplest systems that provide a given nonlinearity to a given accuracy. The result is a method that can engineer both nonlinearities and nonlinear probes, and suggests that both the accuracy and the range of these nonlinearities can be increased by increasing the number of qubits in the auxiliary system. (We note that this work was not, in fact, inspired by nonlinear optics, but by the recent work on ``quantum gadgets", perturbative auxiliary systems, introduced to connect other systems together, for enabling analyses of complexity in quantum computing~\cite{Kempe06}.)

The basic idea is as follows. We couple a mesoscopic resonator, with annihilation operator $a$ and dimensionless position $x = (a + a^\dagger)/\sqrt{2}$, to a low-dimesional auxiliary system via the linear interaction $H_{\mbox{\scriptsize int}}= \mu x V$. Here $V$ is an operator of the auxiliary system, and $\mu$ gives the interaction strength. Interactions proportional to $x$ are straightforward to engineer between mesoscopic resonators and charged systems such as Cooper-pair boxes~\cite{Makhlin01}), polar molecules~\cite{Andre06}, or quantum dots~\cite{Srinivasan07}. How we design the system so that $V$ is simple will be described below. We now choose the auxiliary system so that the separation between its adjacent eigenstates, $\Delta$, is significantly larger than the maximum eigenvalue of the operator $\mu x V$ that will be explored during the evolution of the resonator. The joint Hamiltonian of the two systems may be written as  
\begin{equation}
   H = H_0 + \mu x V  + H_{\mbox{\scriptsize res}} ,
\end{equation}
where $H_0$ is the Hamiltonian of the auxiliary system and $\mu x V$ is a perturbation to this Hamiltonian. Since $x$ commutes with all operators of the auxiliary, we can use standard time-independent perturbation theory to diagonalize $H_{\mbox{\scriptsize aux}} = H_0 + \mu x V$. This gives us the eigenvalues of $H_{\mbox{\scriptsize aux}}$, $E_n$, as a power series in $\lambda = \mu x$. That is, $E_n = \Delta \sum_{m=0}^{\infty} E_n^{(m)} (\mu/\Delta)^m x^m$ for some real numbers $E_n^{(m)}$. If we now place the auxiliary system in one of the eigenvalues of $H_{\mbox{\scriptsize aux}}$, say $E_1$, then the Hamiltonian for the resonator becomes $ H_{\mbox{\scriptsize eng}}   =  \Delta \sum_{m=1}^{\infty}E_n^{(m)} \varepsilon^m x^m + H_{\mbox{\scriptsize res}}$, where $\varepsilon \equiv \mu/\Delta$. By designing $H_0$ and $V$ so as to eliminate specific terms in this sum, we will be able to generate nonlinearities of the form $x^m$ for the resonator. 

For the auxiliary system to act as a probe to measure a nonlinear observable $A$, we must create an effective interaction $H_{\mbox{\scriptsize int}} = \mu A V'$ where $V'$ is an operator of the auxiliary. Information about the resonator operator $A$ can now be obtained by measuring an observable of the auxiliary that does not commute with $V'$ (this is merely the standard von Neumann measurement prescription~\cite{vonNeumann55}). We can engineer a nonlinear interaction  by tailoring the expansions of {\em two} eigenstates of the auxiliary, and placing the auxiliary in the space spanned by these eigenstates. In this subspace, the resulting Hamiltonian has the same form as $H_{\mbox{\scriptsize eng}}$ above, but the numbers $E_n^{(m)}$ become operators in this subspace. By eliminating specific terms in this sum, we can obtain an interaction proportional to a given power of the observable $A$. 

Our task is not merely to obtain an $H_0$ and $V$ for the auxiliary that gives the desired perturbation expansion, but to find those that are simple to implement. To do this we note that a physical interaction $\mu x V$ is usually straightforward to construct (i.e. natural) when $V$ is diagonal in the charge basis of the auxiliary system. We start be specifying a diagonal auxiliary Hamiltonian $\tilde{H}_0$, and perform the perturbation calculation to determine the required interaction operator $\tilde{V}$. We then find the unitary, $U$, that diagonalizes $\tilde{V}$, and this gives us our diagonal $V$ ($V \equiv U\tilde{V}U^\dagger$). The Hamiltonian for the auxiliary system is then given by $H_0 = U\tilde{H}_0U^\dagger$. This is not diagonal in the charge basis, and is the Hamiltonian that we must engineer for the auxiliary system. For a given problem we will search for those auxiliary Hamiltonians that are the simplest to physically construct. 

In what follows we will consider using auxiliary systems of dimension two, three and four to generate nonlinearities. Note that to generate a nonlinearity $x^n$, we should eliminate from the perturbation expansion all terms of order less than $n$, since these will otherwise dominate. It thus makes sense to choose the diagonal elements of $\tilde{V}$ to be zero to eliminate all first-order terms. It is also desirable to eliminate as many terms as possible whose order is higher that $n$. While these terms decrease with increasing order, to give an accurate rendering of $x^n$ we will eliminate the term of order $n+1$, and minimize the higher-order terms.  To proceed we need the expressions for the coefficients of the various orders, $E_n^{(m)}$, of the perturbation expansions of the eigenvalues. We now give these up to fourth-order, where each expression for $E_n^{(m)}$ has been simplified by the condition that $E_n^{(j)}=0$ for $1\leq j<m$. With this simplification the coefficients are 
\begin{eqnarray}
   E^{(2)}_n \! & \! = \! & \! \sum_{m\not= n} \! |\tilde{V}_{mn}|^2 / \Delta_{mn} , 
   \label{E2} \\
   E^{(3)}_n \! & \! = \! & \! \sum_{m,l\not= n} \!\! \tilde{V}_{nm}\tilde{V}_{ml}\tilde{V}_{ln} / (\Delta_{mn}\Delta_{ln}) , 
   \label{E3} \\
   E^{(4)}_n \! & \! = \! & \! \sum_{m,l,j\not= n} \!\!\! \frac{\tilde{V}_{nm}\tilde{V}_{ml}\tilde{V}_{lj}\tilde{V}_{jn}}{\Delta_{mn}\Delta_{ln}\Delta_{jn}}  -  \sum_{m,l\not= n} \!\!\!  \frac{|\tilde{V}_{mn}|^2 |\tilde{V}_{ln}|^2}{\Delta_{mn}\Delta_{ln}^2} , 
   \label{E4}
\end{eqnarray}
where the $\tilde{V}_{nm}$ are the matrix elements of $\tilde{V}$. Higher-order terms are significantly more complex, but are easily calculated numerically. 

To begin we consider a single auxiliary qubit. For a single qubit we are essentially restricted to $H_0 = (\Delta/2) \sigma_x$ and $V = \sigma_z$. The resulting expansion has only terms of even order. To fifth order the effective Hamiltonian is  
\begin{equation}
 H_{\mbox{\scriptsize eff}} = \Delta \sigma_x' \left[ (1/2) - 2 \varepsilon^2  x^2 + 
                                                 6 \varepsilon^4  x^4\right]  + H_{\mbox{\scriptsize res}} ,  
  \label{Heff2D}
\end{equation} 
where $\sigma_x' \equiv  (1-2\varepsilon^2 + 6 \varepsilon^4) \sigma_x + (2\varepsilon - 4\varepsilon^3) \sigma_z$~\footnote{We denote effective Hamiltonians for the resonator alone by $H_{\mbox{\scriptsize eng}}$, and those for the joint system as  $H_{\mbox{\scriptsize eff}}$.}. We cannot engineer the nonlinear potential $x^4$ by itself with a single qubit, since we do not have the freedom to eliminate the second-order term without removing the fourth-order term. However, we see that a single qubit generates the interaction $\sigma_x' x^2$ accurate to third-order in $\varepsilon$. This allows us to measure $x^2$ by using a single-electron transistor to measure $\sigma_z$ for the qubit~\cite{Jacobs08}. It was shown recently that this nonlinear measurement is useful, as it will generate mesoscopic-superpositon states of the resonator directly from a thermal state~\cite{Jacobs08x}. 

In the above case, by choosing the resonator frequency, $\omega$, to be much larger than the strength of the interaction, $\mu = \Delta\varepsilon$, and making the rotating-wave approximation, we obtain the familar energy-coupling $\sigma_x' a^\dagger a$~\cite{Irish03}. If the resonator frequency is much larger than the interaction strength, and we do not wish the $x^2$ coupling to degenerate to the energy coupling, then we can modulate the interaction strength $\mu$ at a frequency, $\nu$, close to $\omega$. This neat trick effectively brings the resonator frequency down to $\omega-\nu$ from the point of view of the coupling~\cite{Jacobs07b}.

The above analysis shows that to generate $x^3$, the auxiliary system must have at least three levels. We now examine what can be achieved with a three-level (qutrit) auxiliary. We take the energy levels of $\tilde{H}_0$ to be evenly separated by $\Delta$, so that $E_n^{(0)} = n\Delta$, $n=0,1,2$, and examine the perturbation expansion for the central level, $E_1$. From Eq.(\ref{E2}) we find that we can eliminate the second-order term by choosing $\tilde{V}$ so that $|\tilde{V}_{12}|^2 = |\tilde{V}_{01}|^2$. Conveniently, for a qutrit, this also removes the fourth-order term. Thus, placing the qutrit in the eigenstate corresponding to $E_1$, this single condition generates the resonator Hamiltonian 
\begin{equation}
H_{\mbox{\scriptsize eng}} = \left( 2 \Delta \mbox{Re}[  \tilde{V}_{01}\tilde{V}_{12}\tilde{V}_{20}] \varepsilon^3 \right) x^3  +  H_{\mbox{\scriptsize res}} , 
\end{equation} 
accurate to fourth order in $\epsilon$. Note that the condition $|\tilde{V}_{12}|^2 = |\tilde{V}_{01}|^2$ gives us considerable scope in choosing the physical interaction operator $\tilde{V}$. In particular, we have complete freedom in choosing the diagonal elements of $V$. On the contrary, all the elements of $H_0$, the physical Hamiltonian for the qutrit, are non-zero. This means that one must couple all three charge states of the qutrit. If the qutrit is a polar molecule or quantum dot, then this could be achieved using three lasers tuned respectively to the three transitions. 

An auxiliary qutrit does not have enough degrees of freedom to generate the nonlinear potential $x^4$ (while removing the second and third-order terms), or indeed to realize a probe for measuring $x^3$. For these tasks one must use an auxiliary with at least four-levels (two qubits). We now turn to this case, and consider first engineering an $x^4$ potential. Once again we take the energy levels of $\tilde{H}_0$ to be $E_n^{(0)} = n\Delta$, and examine the expansion of $E_1$. Using Eqs.(\ref{E2}) and (\ref{E3}), we find that the condition required to eliminate the second-order term is $|\tilde{V}_{13}|^2  =  2 ( |\tilde{V}_{01}|^2 - |\tilde{V}_{12}|^2 )$, and to eliminate the third-order term is 
$\mbox{Re} \left[ 2 \tilde{V}_{10}\tilde{V}_{02} \tilde{V}_{21} + \tilde{V}_{10}\tilde{V}_{03} \tilde{V}_{31} - \tilde{V}_{12}\tilde{V}_{23} \tilde{V}_{31}\right] = 0$ . These two conditions also remove the fifth-order term. 

To find practical auxiliary systems, we perform a numerical search over all $\tilde{V}$ that satisfy these conditions, minimizing the distance between the resulting auxiliary Hamiltonian, $H_0$, and those that can be readily engineered. So long as the two conditions are enforced, the optimization is not difficult and gives many solutions. A nice example is the two-qubit Hamiltonian 
\begin{equation}
  H_0 =  \Delta \left[ a \sigma_z^{(1)} \sigma_z^{(2)}  + b \sigma_x^{(2)} - c  \sigma_x^{(1)} \sigma_x^{(2)}  \right] , 
  \label{Hx4}
\end{equation} 
where $a = 0.914$, $b = 0.405$ and $c = 0.5$. The interaction terms in this Hamiltonian can be engineered between superconducting qubits using, for example, methods devised by Mooij {\em et al.}~\cite{Mooij99}. The required interaction with the target resonator is  
\begin{equation} 
  V =   \mu \left[ f \sigma_z^{(1)} + g \sigma_z^{(2)} \right] x , 
  \label{Hintx4} 
\end{equation}
where $f = -1.823$ and $g = -1.382$. This auxiliary system generates an expansion for $E_1$ in which all the terms of odd-order are eliminated (at least up to seventh order) and the coefficients of the even terms are, $E_1^{(2)} = 2\times 10^{-4}$, $E_1^{(4)}  = -1$ and $E_1^{(6)}  = -3.99$. It thus generates an $x^4$ potential accurate to fifth order in $\varepsilon$. 

The above auxiliary system also allows us to generate the Kerr (or $\chi^{(3)}$) nonlinearity, whose Hamiltonian is $H_{\mbox{\scriptsize Kerr}} \propto (a^\dagger a)^2$.  To do this we use exactly the same adiabatic elimination process that turns the interaction $\sigma_x x^2$ into $\sigma_x a^\dagger a$: we choose the resonator frequency to be much larger than the strength of the $x^4$ potential and make the rotating wave approximation. This eliminates all terms in $x^4$ that are not powers of $a^\dagger a$, transforming $x^4$ into $(3/2)(a^\dagger a)^2$, and thus $\chi^{(3)}$.  

Now consider engineering a probe for measuring $x^4$ (or equivalently ($a^\dagger a)^2$). To do this we must tailor the expansions of two eigenvalues. If we choose $E_1$ and $E_2$, then this merely involves using the conditions above for tailoring $E_1$, and adding to them the equivalent conditions for $E_2$ (these can be obtained from the former by symmetry). Performing a numerical search to find $H_0$ and $V$, our results suggest that there is only one solution, being  
\begin{equation}
  H_0 =   \Delta \left[  \sigma_x^{(1)} +  (1/2) \sigma_x^{(2)} \right] , 
  \label{hammeasx4}
\end{equation}
with the interaction operator $V =  \mu [ f \sigma_z^{(1)} + g \sigma_z^{(2)}] x$, where $f = 1.682$ and $g = 1.189$. This Hamiltonian is even simpler than the one presented in Eq.(\ref{Hx4}), as no coupling is required between the two qubits. This Hamiltonian can, of course, also be used to engineer $x^4$ and $\chi^{(3)}$ nonlinearities by placing the auxiliary in the eigenstate with eigenvalue $E_1$ (or $E_2$). This configuration eliminates all the odd terms in the expansion for $E_1$ and $E_2$ up to seventh order, and generates the even terms $E_2^{(2)} = - E_1^{(2)} = 8\times 10^{-4}$, $E_2^{(4)} = - E_1^{(4)} = 1$ and $E_2^{(6)} = - E_1^{(6)} = -4.24$. Denoting the eigenstates of the auxiliary with eigenvalues $E_1$ and $E_2$ as $|1\rangle$ and $|2\rangle$, this auxiliary system generates the effective interaction $H_{\mbox{\scriptsize eff}} =  \Delta \varepsilon^4 Y x^4$, where $Y = |1\rangle \langle 1| - |2\rangle\langle 2|$. 

To show that our method can be realized with current technology, we consider implementing an $x^4$ nonlinearity using the configuration given in Eq.(\ref{hammeasx4}). This is achieved by coupling the resonator to two Cooper-pair boxes 
(CPBs) via the usual capacitative interaction~\cite{Armour02}. Operating each CPB at its degeneracy point  gives exactly the right Hamiltonian; the Hamiltonian for the $j^{\mbox{\scriptsize\textit{th}}}$ CPB is $H_j = \hbar\omega^{\mbox{\scriptsize J}}_j \sigma_x$, where $\hbar\omega^{\mbox{\scriptsize J}}_j$ is the (tunable) Josephson energy~\cite{Makhlin01}. Reasonable values are $(\omega^{\mbox{\scriptsize J}}_1,\omega^{\mbox{\scriptsize J}}_2) = (1,1/2) \times 10^{9} \mbox{s}^{-1}$~\cite{Majer07, Houck07}. Scaling the resonator's position so that $x = (a + a^\dagger)/\sqrt{2}$, current interaction strengths for superconducting resonators are $\mu \sim 10^{8}$~\cite{Houck07}, and this is also realistic for nanomechanical resonators~\cite{Armour02}. With these parameters, and placing the qubits in the eigenstate of $E_{2}$ (achieved using simple single-qubit operations) the engineered Hamiltonian is 
\begin{equation} 
  H_{\mbox{\scriptsize eng}} =  \hbar \kappa \left[  x^4  -  0.042\, x^6 + \mathcal{O}(\varepsilon^8) \right]  + H_{\mbox{\scriptsize res}}, 
  \label{eq::finham}
\end{equation} 
with $\kappa = 10^5 \mbox{ s}^{-1}$. This is a giant $x^4$ nonlinearity as
advertised, albeit with a (relatively) small additional $x^6$ term. With a 
resonator frequency of $100~\mbox{MHz}$, this $x^4$ term is to excellent 
approximation a $\chi^{(3)}$ nonlinearity, given by the Hamiltonian 
$H_{\mbox{\scriptsize eng}} =  - (3/2) \hbar \kappa (a^\dagger a)^2$. As a 
benchmark for the strength of this nonlinearity, one can use the time it takes to create a mesoscopic-superposition of two coherent states, starting from a coherent state. This time is conveniently independent of the amplitude of the initial coherent state. The nonlinearity above performs this task in $\tau = \pi /(3\kappa) = 10 \,\mu \mbox{s}$~\cite{Jacobs07x}. (To compare, resonator damping times are $\sim10 \, \mbox{ms}$ for mechanical~\cite{LaHaye09},  and $\sim10 \, \mu\mbox{s}$ for superconducting~\cite{Houck07}.)

We now consider the robustness of the method to errors in the auxiliary Hamiltonian. To do this we add an independent Gaussian error with standard deviation $0.01\omega^{\mbox{\scriptsize J}}_1$ (errors of $\sim 1$\%) to the three degrees of freedom of each CPB Hamiltonian, and calculate the resulting perturbation expansion for $E_2$. We then calculate the standard deviation of the induced change in each term in the expansion of $E_2$ by averaging over many samples. Denoting the standard deviation of the change in the coefficient $E_2^{(m)}$ as $\Sigma_{(m)}$, we find that these are very small: $(\Sigma_{(1)},\ldots,\Sigma_{(7)}) = 10^{-3}\times (0.2, 0.3, 0.8, 1.3, 3.3, 5.8, 15.3)$. The method is thus quite robust to errors. This calculation also provides an estimate of the errors induced by dephasing, since this noise can be modeled as fluctuations in the Hamiltonian. Errors from dephasing should be significantly smaller than those calculated above, since the best dephasing rates for CPB's are currently $10^{6} \, \mbox{s}^{-1} \approx 10^{-3} \omega^{\mbox{\scriptsize J}}_1$~\cite{Lehnert03}. We can also estimate the effects of noise due to decay of the upper levels of the CPBs in the same way. The broadening of the energy levels for each CPB is given by the non-Hermitian Hamiltonian $H = -i\hbar (\gamma/2) \sigma_+\sigma_-$, where $\gamma \approx 10^{-3} \omega^{\mbox{\scriptsize J}}_1$ is the decay rate. Including this Hamiltonian and re-calculating the perturbation expansion reveals that the changes induced in each coefficient are no greater than $10^{-3}$.  

Returning to Eq.(\ref{eq::finham}), we can reduce the relative size of the sixth-order correction by increasing $\Delta$ with respect to $\mu$, which is quite practical. However, this will also reduce the strength of the $x^4$ nonlinearity. If we want to both increase the $x^4$ term and reduce the $x^6$ term, then we need to reduce the coefficient of the sixth-order term in the perturbation expansion. Our numerical search  indicates that there is a limit to how small one can make the sixth-order term with a 4-level auxiliary. We expect that using an auxiliary with 3 qubits (8 levels) will allow elimination of the sixth-order term completely.      

As useful as the results we have presented here are for engineering mesoscopic systems, this technique appears to promise even greater potential with the use of larger auxiliary systems (e.g. three qubits). Realizing this potential will require exploring the complex landscape that maps Hamiltonians to perturbation expansions, and this is an interesting question for future work. 

{\em Acknowledgements:} We thank Salman Habib for hospitality at the QUEST
workshop in Santa Fe that initiated the exchange of ideas leading to this
work. KJ is supported by the Hearne Institute for Theoretical Physics, ARO
and IARPA. AJL's work was supported by the Center for Advanced Studies at
UNM, and the NSF under contracts PHY-0555573, PHY-0653596, and CCF-0829944.
AJL's work is currently supported by Sandia National Laboratories, a
multi-program laboratory operated by Sandia Corporation, a Lockheed-Martin
Company, for the U.S.\ Department of Energy under Contract No.\
DE-AC04-94AL85000.



\end{document}